# Local texture of three-stage CVD SiC fibre by precession electron diffraction (PED) and XRD

B. Huang*[1], Y. Q. Yang[1], M. H. Li[1], Y. X. Chen[1], X. Luo[1], M. S. Fu[1], Y. Chen[1] and Xierong Zeng[2]

SiC fibre with the transverse isotropic properties is very important to it reinforced metal matrix composites. In this paper, local texture of the CVD SiC fibre was investigated by means of X-ray diffraction (XRD) and precession electron diffraction (PED) on transmission electron microscopy (TEM). The result from XRD is in agreement with the result obtained from PED. And the result shown that at the first stage of deposition, the preferred direction of SiC grains is almost random and the distribution of grain size is scattered. At the second and third stages of deposition, there are two kinds of texture in SiC fibre, that is, (110)<111> and (110)<115>. Furthermore, the grain size at the second and third stages is about 200 nm and it is lower at the third stage than at the second stage because of the lower temperature at the third stage. The [110] preferred direction along axial direction for SiC fibre is beneficial to the axial tensile strength.

## Introduction

Owing to its excellent mechanical and physical properties, such as high tensile strength and stiffness, high oxidation resistance and good compatibility with a variety of metal materials, continuous SiC fibre manufactured by chemical vapour deposition (CVD) can be used to fabricate the metal matrix composite, which is the most promising candidate for the high temperature application, especially in advanced aircraft and aerospace vehicles.[1–4]

It is well known that the properties of SiC fibre, depending on its microstructure, will influence the properties of the composite. For example, the transverse isotropic properties of SiC fibre result in different properties, especially the tensile strength, between the longitudinal and transverse directions for the composites. This point is dramatically relevant to the texture in SiC fibre. The microstructure of SCS-6 SiC fibre has been studied by Ning and Pirouz.[5] They reported that the elongated grains in SiC-3 and SiC-4 areas (its width is about 40 μm) have a highly preferential growth direction along <111> direction of SiC parallel to the radial direction of the fibre. On the other hand, Cheng et al.[6] investigated the microstructure of sigma 1140+ SiC fibre by using electron diffraction. And the results show that the grains are strongly textured with their <111> directions roughly parallel to the fibre radius. In addition, Radmilovic et al.[7] studied the evolution of the morphology and the texture of 3-C SiC films grown on the Si(100) substrates by CVD. The results indicated the film exhibited a columnar grain structure with a strong <111> fibre texture and a high density of stacking faults and twins. In our previous work, the texture of CVD SiC fibre has been studied by analysing a single profile from X-ray diffraction (XRD). We found that the deposited SiC of the fibre has a <111> preferred orientation. As illustrated above, the high preferential growth <111> direction parallels to the radial direction of SiC fibre. However, generally, the effect of properties of SiC fibre on the axial direction is very important to the reinforced composite. Therefore, the texture of SiC fibre along the axial direction needs to be further research. In addition, as shown in previous studies,[8–11] there are two or three concentric sublayers along the radial direction because of the two or three deposition stages during the CVD process. Moreover, the microstructure of SiC fibre, such as the grain size, is different for each concentric sublayer. Then, the question is whether the local texture for each sublayer is different or whether there are some relationships among the local textures for these sublayers.

Precession electron diffraction (PED) allows for diffraction pattern collection under quasi-kinematical conditions.[12] Diffraction patterns are collected, while the transmission electron microscopy (TEM) electron beam is precessing on a core surface; in this way, only a few reflection are simultaneously excited and, therefore, dynamical effects are strongly reduced.[13–15] The combination of PED with fast electron diffraction acquisition and pattern matching software techniques can be used for the high magnification ultrafast mapping of variable crystal orientations and phases, similarly to what is achieved with the electron backscattered diffraction

[1] State Key Lab of Solidification Processing, Northwestern Polytechnical University, Xi'an 710072, China
[2] College of Materials Science and Engineering, Shenzhen University, Shenzhen 518060, China

*Corresponding author, email huangbin@nwpu.edu.cn

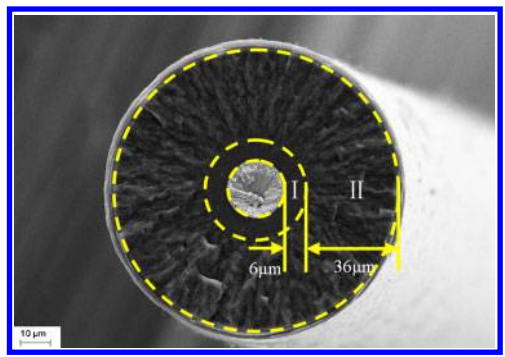

1 Morphology of fracture surface for SiC fibre

technique in scanning electron microscope (SEM) at lower magnifications and longer acquisition times.[16–23]

In this work, we will study the local texture in the SiC fibre by using XRD and PED and discuss its evolution along the radial direction.

## Materials and experimental

### Materials

The continuous SiC fibre studied in this work was fabricated by three-stage CVD on a tungsten filament heated by direct current in a $CH_3SiCl_3–H_2$ system.[8] Tungsten filament (purity ≥ 99%) with a diameter of 17·6 μm was passed through the three-stage horizontal reactor. Mercury electrodes at each stage end allowed the direct current to pass through the filament. The pressure of each stage was 0·1 MPa. The tungsten filament was cleaned in hydrogen at the first stage and then a thin SiC layer was grown on it at the second stage. At the third stage, the second layer of SiC was deposited on the SiC (deposited during the preceding stage).

For the sake of convenience, in this paper, local texture measurements were carried out on the sample of SiC fibre reinforced Ti–6Al–4V matrix composite. The composite was fabricated by using foil–fibre–foil method. The TEM sample cut from the as processed composite bulk was ground using abrasive paper, followed by dimpling on the Gatan 582 dimple grinder. The final milling was performed on the Precision Ion Polishing System from Gatan.

### Experimental

The morphology of SiC fibre was observed on ZEISS SUPRA 55 field emission scanning electron microscope (SEM). Then, both structures in cross-section and longitudinal section of SiC fibre were investigated by using X-ray diffractometer from PANalytical. When the structure in longitudinal section was studied, the rotation axis of the goniometer is parallel to the axial direction of SiC fibres. It means that the crystal planes, which are parallel to the radial direction of SiC fibres, will be investigated. On the contrary, the rotation axis of the goniometer is perpendicular to the axial direction of SiC fibres in cross-section measurement. And it is mean that the crystal planes, which are parallel to the axial direction of SiC fibres, will be investigated in cross-section measurement. Based on the morphology and structure information, the local texture measurements in SiC fibre were carried out on FEI Tecnai F30 field emission transmission electron microscope (TEM), which is equipped with PED system from NanoMEGAS.

## Results and discussion

### Morphology and microstructure of SiC fibre by SEM

The morphology of fracture surface for SiC fibre is shown in Fig. 1. It can be seen that the SiC fibre with a diameter of about 100 mm includes tungsten core (about 18 mm thick), a W/SiC interfacial reaction zone and deposited SiC layers. Furthermore, according to the morphology of deposited SiC, it can be divided into two sublayers: layer I with a neat fracture surface (about 6 μm) and layer II with a rough fracture surface (about 36 μm).

### Microstructure and texture of SiC fibre by XRD

Figure 2 shows the XRD patterns of the SiC fibre in longitudinal section and cross-section. It can be seen that the fibre is mainly composed of β-SiC. The preferred orientation of polycrystalline SiC layer can be described by the texture coefficient (TC) of the Harris method[24]

$$TC = \frac{I/I_0}{(1/n)\sum_n (I/I_0)} \quad (1)$$

where $I$ is the measured intensity, $I_0$ is the standard intensity of International Centre for Diffraction Date and $n$ is the number of reflections. The calculated TCs of the crystal planes in the cross-section and longitudinal section are shown in Table 1. It is obvious that TCs of (1 1 1) and (2 2 2) reflections are biggest in longitudinal section. Additionally, TC of (1 1 5)/(3 3 3) reflection is also high in longitudinal section. Because the plane distances of (1 1 5) and (3 3 3) are equal to each other at the cubic crystalline, the two planes cannot be discriminated in the same XRD pattern. Therefore, the deposited SiC has (111) and maybe has (115) preferred orientation in longitudinal section. On the other hand, for the cross-section measurement, TC of (110) reflection is dramatically bigger than those of others. Therefore, it can be concluded that (110)<111> texture exists in SiC fibre and (110)<115> texture may exist in SiC fibre. Additionally, the diffraction peaks from W core and interfacial reaction productions between W core and deposited SiC, such as $W_2C$ in Fig. 2b, also can be seen in Fig. 2. According to the studies by Guo[25] and Cheng,[6] there is a reaction zone between SiC and W core. The details of interfacial reaction products are not certain; however, $W_2C$ presents at the both interface in the works by Guo and Cheng. In addition, the reference intensity ratio of $W_2C$, which represents the reflection ability of the product to X-ray and can be found from International Centre for Diffraction Date database, is the highest among the interfacial reaction products. Therefore, only the diffraction peaks from $W_2C$ can be seen in Fig. 2b.

Table 1 Calculated TCs of crystal planes in cross-section and longitudinal section

|  | 111 | 200 | 220 | 311 | 222 | 400 | 331 | 420 | 422 | 511 |
|---|---|---|---|---|---|---|---|---|---|---|
| Cross-section | 0·18 | 0·07 | 3·21 | 0·54 | … | … | … | … | … | … |
| Longitudinal section | 2·95 | 0·32 | 0·50 | 0·56 | 2·67 | 0·36 | 0·42 | 0·87 | 0·39 | 0·97 |

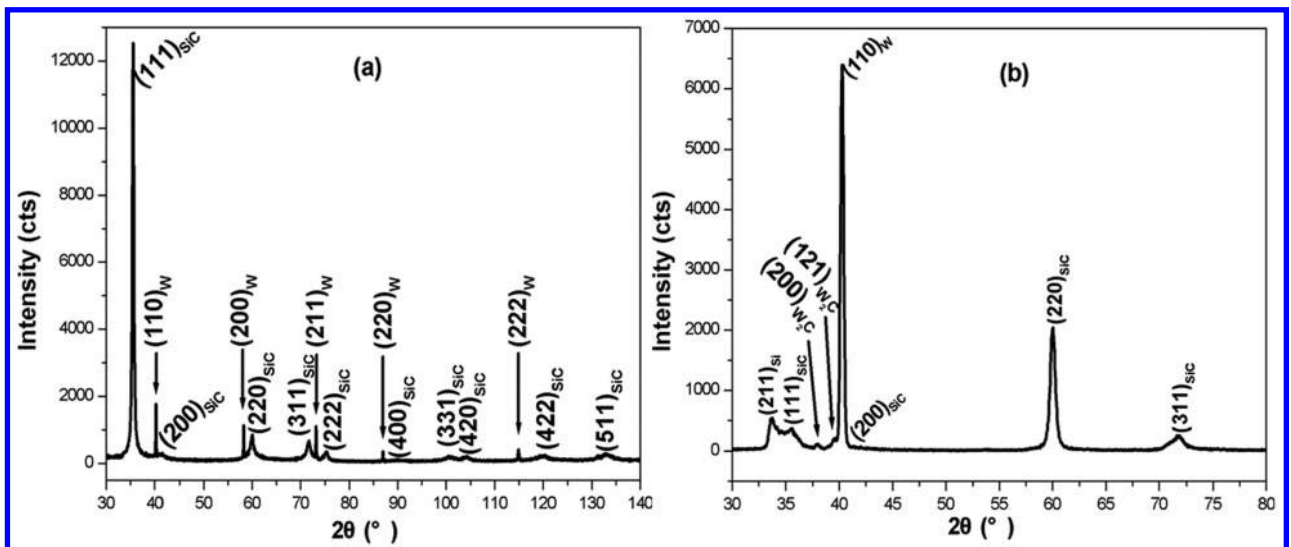

**2** Diffraction patterns for SiC fibre in *a* longitudinal section and *b* cross-section

## Local texture of SiC fibre by PED

According to the fabrication process and morphology of SiC fibre, three areas (A, B and C in Fig. 3) will be selected to study the local texture. A is located at layer I (as indicated in Fig. 1) and B and C are located at layer II (as indicated in Fig. 1). The coordinate system for the local texture measurement was also illustrated in Fig. 3. In this system, $X$ axis is parallel to radial direction of SiC fibre, which is rolling direction (RD). $Y$ axis is normal to radial direction of SiC fibre, which is transverse direction (TD). $Z$ axis is parallel to longitudinal direction of SiC fibre, which is normal direction (ND).

### Local texture for A area

The local texture for the grains at area A of SiC fibre is shown in Fig. 4. Figure 4*a* and *b* are the bright filed image from TEM and orientation distribution using inverse pole figure along $X$ direction respectively. It is obvious that the column grains appear at this area and the distribution of diameters for the SiC grains is scattered. For example, the largest diameter of SiC grain is up to 400 nm and the smallest is less than 50 nm. Figure 4*c* and *d* are pole figure of (110) and orientation distribution function (ODF) ($\Phi=0°$) for the grains at this area of SiC fibre respectively. It can be seen from Fig. 4*c* and *d*, especially the ODF result, there are several preferred orientations for the grains at this area. In other words, the preferred orientation for these grains at area A is not clear. That is, to some extent, during the initial stage, the orientation for these grains of SiC fibre is random.

### Local texture for B area

Figure 5 shows the local texture for the grains at area B of SiC fibre. The bright filed image from TEM and orientation distribution using inverse pole figure along X direction are shown in Fig. 5*a* and *b* respectively. It can be seen from Fig. 5*a* and *b* that the column grains appear at this area and its width is about 200 nm. Compared with it at area A, the width of column grains at B area is smaller. Figure 5*c* and *d* shows pole figure of (110) and ODF ($\Phi=0°$) for the grains at this area of SiC fibre respectively. According to the pole figure of (110) for SiC, as shown in Fig. 5*c*, generally, there are only two preferred orientations, (110)<111> and (110)<115>, for these grains at area B of SiC fibre. That is, during the initial second stage, the orientations of the grains at this area are grouped at (110)<111> and (110)<115>. This result is in good agreement with those obtained from XRD. Meanwhile, it can also be concluded from Fig. 5*d* that preferred orientation of grains at this area of SiC fibre is very clear. Generally, in pole figure and ODF figure, the density contour represents the intensity of textures. For example, in Fig. 5*d*, the densities of the two textures at B area, (110)<111> and (110)<115>, are 22·0 and 25·1 respectively. As a result, the intensity of texture (110)<115> is slightly higher than that of texture (110)<111> for the grains at area B of SiC fibre.

### Local texture for C area

Figure 6*a* and *b* are the bright filed image from TEM and orientation distribution along $X$ direction using inverse pole figure for C area of SiC fibre respectively. It can be observed in Fig. 6*a* and *b*, similarly to that at B area, that the column grains with 200 nm width appear at this area. Figure 6*c* and *d* shows pole figure of (110) and ODF ($\Phi=0°$) for the grains at this area of SiC fibre respectively. It can be concluded from the pole figure of (110) that the preferred orientation for the grains at this area of SiC fibre is clear. For the texture at B area, there are only two preferred orientations, (110)<111> and (110)<115>, for these grains at C area of SiC fibre. However, comparing with the orientation distribution at B, the intensities of the textures at C area is lower. That is, during the post-second stage, although the orientations for these grains at C area

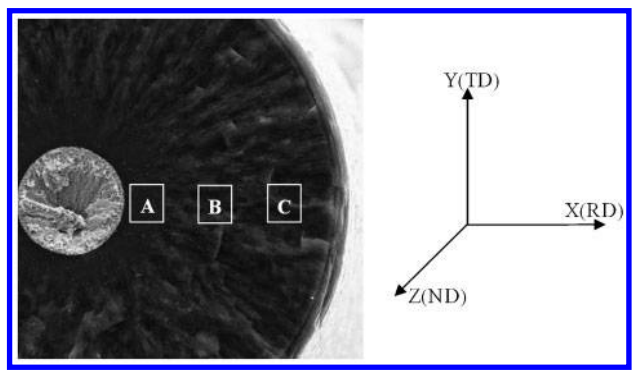

**3** Selected three areas (A, B and C) on cross-section of SiC fibre

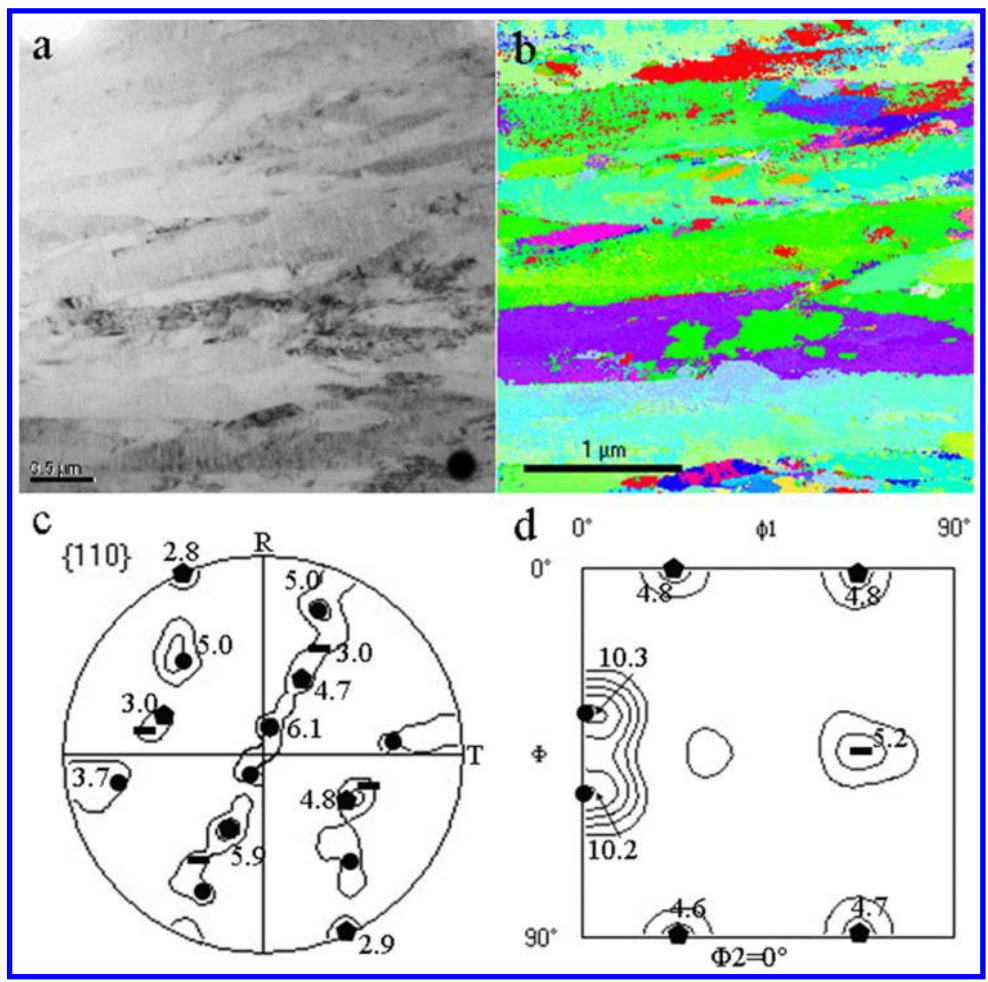

*a* bright field image from TEM; *b* orientation distribution along *X* direction using inverse pole figure; *c* pole figure of (110); *d* orientation distribution function (ODF) for A area

**4 Local texture for grains of SiC at A area**

are still preferred at (110)<111> and (110)<115>, these intensities decrease. The ODF data also verify the result from pole figure of (110). The preferred orientation of grains at C area of SiC fibre is very clear from ODF. In addition, the densities of the two textures at C area, (110)<111> and (110)<115>, are 11·1 and 16·8 respectively. In other words, the intensity of texture (110)<115> is higher than that of texture (110)<111> for the grains at area C of SiC fibre. Furthermore, compared with the grains at area B, the orientations of more grains prefer to (110)<115> at area C.

## Discussion

### Texture and grain growth of SiC fibre during fabrication process

According to the results from XRD and TEM PED, two textures (110)<111> and (110)<115> exist in SiC fibre. Furthermore, the grains with preferred orientation are mostly located at the outer deposited SiC area during the second fabrication stage. On the contrary, during the first fabrication stage, the orientation of SiC fibre is almost random.

At the initial period of the first fabrication stage, Si and C atoms deposit on the W core. Because of the high temperature and the heterogeneous nucleation, many small SiC grains with the different direction formed at the surface of W core, even there is some amorphous layer near the SiC/W interface. With increasing deposition time, these grains will grow up. As shown above, at the region adjacent to the W core (or SiC/W interface), the direction of these grains is almost random and the distribution of grain size is also scattered. At the initial period of the second deposition stage, SiC will be deposited on the original SiC grains formed at the first stage. Because the energy of {111} plane for face core cubic structure is the lowest,[26] the grains with <111> directions grow up slowly along the radial direction of SiC fibre. It means the grains with <111> directions spread normal to the radial direction of SiC fibre. On the contrary, the grains with the non-preferential directions grow up quickly along the radial direction of SiC fibre because of higher energy crystal planes. This will result in the large grains with <111> directions and small grains with the non-preferential directions. At the post-period of the second deposition stage, SiC will be deposited on the original SiC grains formed at the first stage. Therefore, the SiC grains will nuclear and grow up on the previous SiC grains. Compared with the heterogeneous nucleation at the initial period of the first fabrication stage, homogeneous nucleation is primary at this time. Because the energy barrier between the new nuclei and the previous SiC grain is low,[27] the preferential orientations of these grains will be close to that of grains formed at the first stage. As mentioned above, the grains with <111> direction along radial

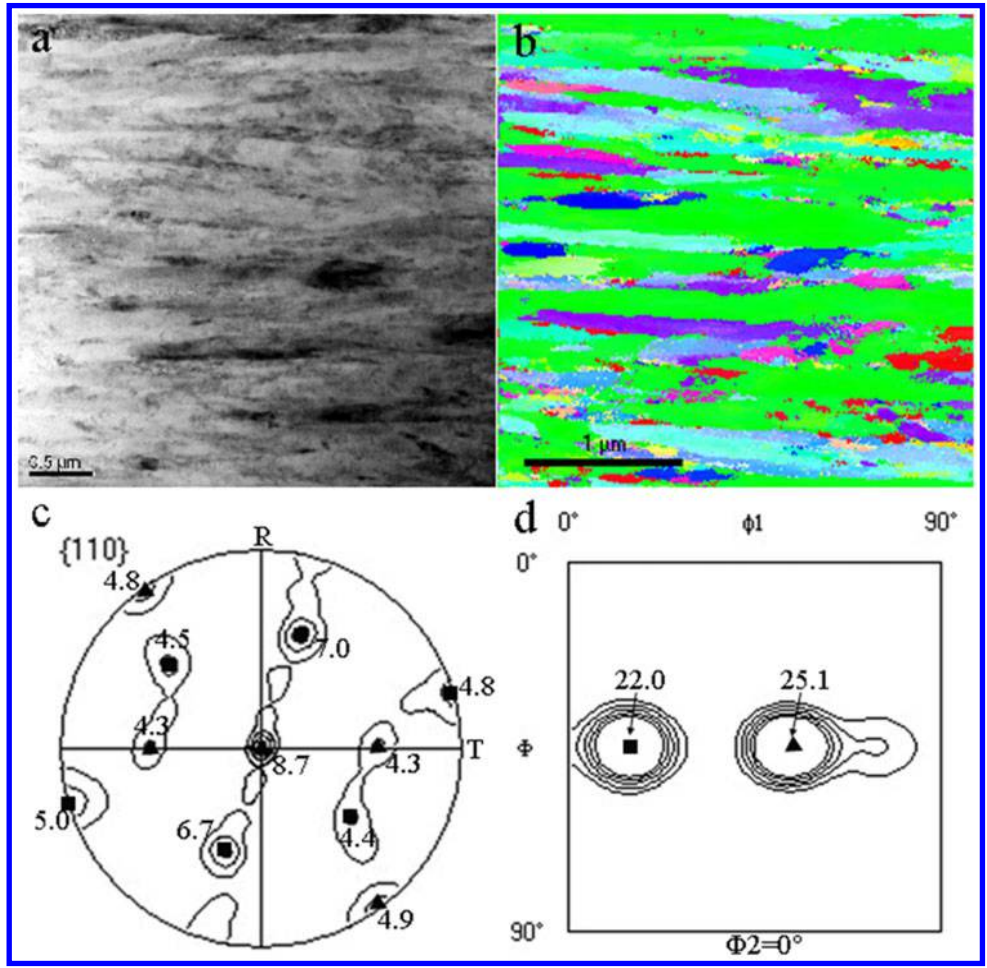

*a* bright field image from TEM; *b* orientation distribution along X direction using inverse pole figure; *c* pole figure of (110); *d* orientation distribution function (ODF) for B area
5 Local texture for grains of SiC at B area direction of SiC fibre will grow up slowly. Finally, the directions of grains are most grouped at <111>. However, the growth speed of the grains at the second stage is lower than that at the first stage due to the lower temperature resulting from the larger diameter of SiC fibre at the second stage. Therefore, at the initial period of the second stage, the width of column grains is lower than that at the first stage and preferential directions of the grains are most grouped at <111> along radial direction. At the post-period of the second stage, there is no noticeable change for the preferential direction of the grains. Only the width of column grains deceases slightly because the temperature decreases.

## Preferred orientation for SiC grains along normal (axial) direction

Generally, the axial tensile strength is very important and is used to judge the quality of SiC fibre with transverse isotropic properties. For example, the axial tensile strength of SCS-6 SiC fibre is up to more 4000 MPa. Furthermore, it is well known that the tensile strength of the brittle materials strongly depend upon its elastic constant and different crystallographic directions and different elastic constants. As a result, for the textured materials, elastic constant for the preferred direction is very important. The elastic constant for the specific crystal direction [$hkl$] can be calculated by the formulation as follows[28]

$$\frac{1}{E^{[hkl]}} = S_{11} - 2\left[(S_{11} - S_{12}) - \frac{1}{2}S_{44}\right]\Gamma^{[hkl]} \quad (2)$$

$$\Gamma^{[hkl]} = \frac{h^2k^2 + k^2l^2 + l^2h^2}{(h^2 + k^2 + l^2)^2} \quad (3)$$

For the cubic system[29]

$$C_{11} = \frac{S_{11} + S_{12}}{(S_{11} - S_{12})(S_{11} + 2S_{12})} \quad (4)$$

$$C_{12} = \frac{-S_{12}}{(S_{11} - S_{12})(S_{11} + 2S_{12})} \quad (5)$$

$$C_{44} = 1/S_{44} \quad (6)$$

were $E^{[hkl]}$ is the elastic constant for the crystal direction [$hkl$], $S_{ij}$ is the compliance coefficient, $C_{ij}$ is the stiffness coefficient and $\gamma^{[hkl]}$ is orientation coefficient.

The elastic stiffness coefficient of cubic SiC fibre can be referred to the data of the cubic (3C) polycrystal of SiC.[30] According to the reference, the elastic stiffness coefficients are $C_{11} = 352 \cdot 3$ GPa, $C_{12} = 140 \cdot 4$ GPa and $C_{44} = 232 \cdot 9$ GPa. Using these data, the elastic compliance coefficient of SiC can be calculated as follows: $S_{11} = 0 \cdot 003673$ GPa$^{-1}$, $S_{12} = -0 \cdot 001047$ GPa$^{-1}$ and $S_{44} = 0 \cdot 004294$ GPa$^{-1}$. As indicated in the cross-section XRD

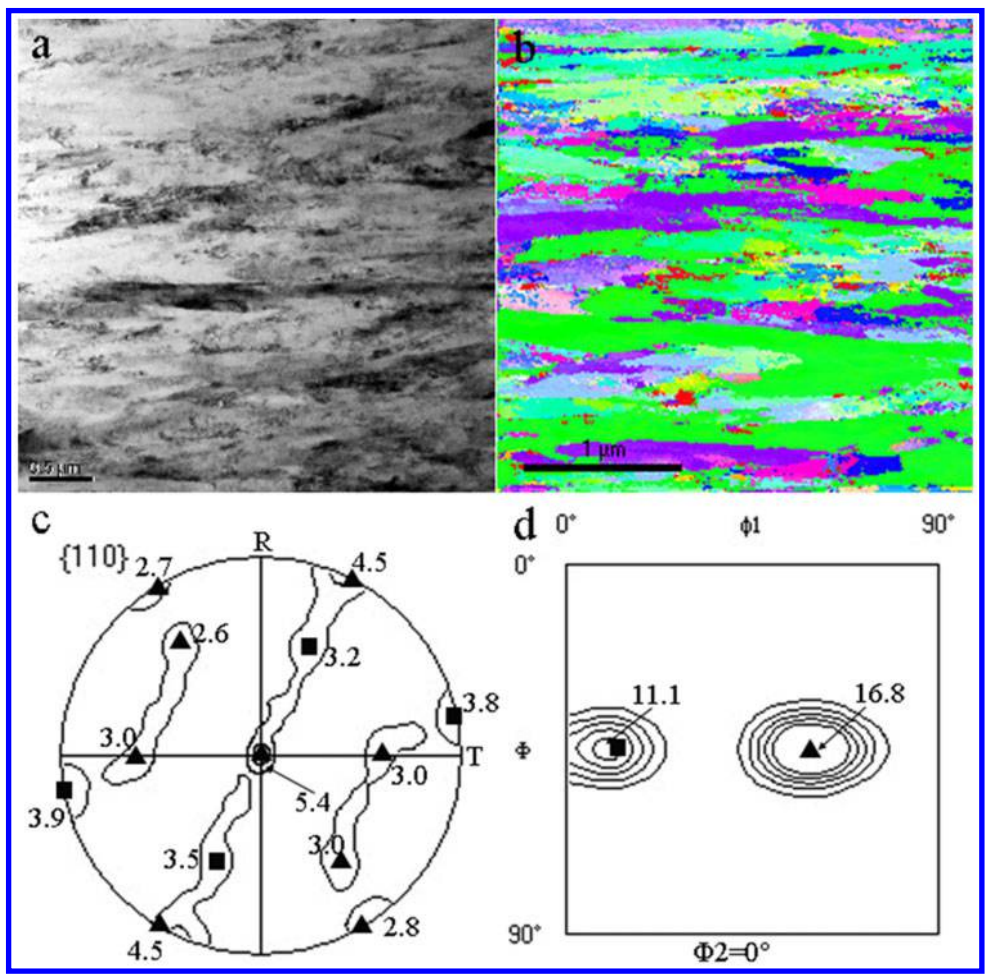

*a* bright field image from TEM; *b* orientation distribution along *X* direction using inverse pole figure; *c* pole figure of (101); *d* orientation distribution function (ODF) for C area

6 Local texture for grains of SiC at C area results, there are four crystal planes, which are (111), (220), (200) and (311), for the investigated range. For a single SiC grain, when its orientation along the radial direction is preferred at <111>, the orientation of this grain along axial direction is only <100> or <110>. Therefore, we calculated the elastic constants for the crystal directions [100] and [110] through equations (2) and (3). The results show that $E^{[100]}=272.3$ GPa and $E^{[110]}=419.0$ GPa. That is, the elastic constant for [110] direction is larger than that for [100] direction. As discussed above, the preferential orientation of these SiC grains is grouped at <110> along axial direction. Therefore, it can be concluded that the preferred direction [110] along axial direction for SiC fibre is beneficial to the axial tensile strength.

## Conclusion

Local texture in SiC fibre was investigated by XRD and PED. The result obtained from two-direction XRD measurement shows that two kinds of texture (110)<111> and (110)<115> exist in SiC fibre. The results from PED measurement verified this point. At the first stage of deposition, the distribution of the grain direction is almost random and the distribution of grain size is scattered. At the second and the third stages of deposition, there are two kinds of texture in SiC fibre, that is, (110)<111> and (110)<115>. Furthermore, the grain size at the second stage is about 200 nm and it is lower at the third stage than that at the second stage because of the lower temperature at the third stage. The preferred direction [110] along axial direction for SiC fibre is beneficial to its axial tensile strength.

## Acknowledgement

The authors gratefully acknowledge the financial supports by the Nature Science Foundation of China (No. 51201135), the NPU Foundation for Fundamental Research (No. NPU-FFR-JC201110) and Foundation - of Shenzhen Key Laboratory of Special Functional Materials (No. T201203).